\documentclass[sigconf]{acmart}

\usepackage{algorithm}
\usepackage[show]{chato-notes}
\usepackage{comment}
\usepackage{cleveref}
\usepackage{algpseudocode}
\usepackage{amsmath}
\usepackage{bbm}
\usepackage{dutchcal}
\usepackage[greek,english]{babel}
\usepackage[show]{chato-notes}
\usepackage{enumerate}
\usepackage{epsfig}
\usepackage{graphicx}
\usepackage{hyperref}
\usepackage[utf8]{inputenc}
\usepackage{booktabs}
\usepackage{adjustbox}
\usepackage{multirow}
\usepackage{subcaption}
\usepackage{url}
\usepackage{xspace}
\usepackage{tikz}
\usepackage[framemethod=TikZ]{mdframed}
\usetikzlibrary{bayesnet}
\usetikzlibrary{positioning}

\hypersetup{
    unicode=false,          %
    pdftoolbar=true,        %
    pdfmenubar=true,        %
    pdffitwindow=false,     %
    pdfstartview={FitH},    %
    pdftitle={My title},    %
    pdfauthor={Author},     %
    pdfsubject={Subject},   %
    pdfcreator={Creator},   %
    pdfproducer={Producer}, %
    pdfnewwindow=true,      %
    colorlinks=true,       %
    linkcolor=black,          %
    citecolor=black,        %
    filecolor=black,      %
    urlcolor=black           %
}

\newcommand{\spara}[1]{\smallskip\noindent{\bf #1}}

\newcommand{\I}{\ensuremath{\mathbb{I}}}

\newcommand{\D}{\ensuremath{\mathbb{D}}}

\newcommand{\p}{\ensuremath{\mathcal{s}}}

\algnewcommand\algorithmicforeach{\textbf{for each}}
\algdef{S}[FOR]{ForEach}[1]{\algorithmicforeach\ #1\ \algorithmicdo}

\renewcommand{\algorithmicrequire}{\textbf{Input:}}
\renewcommand{\algorithmicensure}{\textbf{Output:}}

\newcommand\captionfigdist{-3mm}
\newcommand\figtextdist{-1.5mm}

\newcommand{\Brexit}{\texttt{Brexit}\xspace}
\newcommand{\Referendum}{\texttt{Referendum}\xspace}
\newcommand{\VaxNoVax}{\texttt{VaxNoVax}\xspace}

\newcommand{\squishlist}{
 \begin{list}{$\bullet$}
  {  \setlength{\itemsep}{0pt}
     \setlength{\parsep}{3pt}
     \setlength{\topsep}{3pt}
     \setlength{\partopsep}{0pt}
     \setlength{\leftmargin}{2em}
     \setlength{\labelwidth}{1.5em}
     \setlength{\labelsep}{0.5em}
} }
\newcommand{\squishlisttight}{
 \begin{list}{$\bullet$}
  { \setlength{\itemsep}{0pt}
    \setlength{\parsep}{0pt}
    \setlength{\topsep}{0pt}
    \setlength{\partopsep}{0pt}
    \setlength{\leftmargin}{2em}
    \setlength{\labelwidth}{1.5em}
    \setlength{\labelsep}{0.5em}
} }

\newcommand{\squishdesc}{
 \begin{list}{}
  {  \setlength{\itemsep}{0pt}
     \setlength{\parsep}{3pt}
     \setlength{\topsep}{3pt}
     \setlength{\partopsep}{0pt}
     \setlength{\leftmargin}{1em}
     \setlength{\labelwidth}{1.5em}
     \setlength{\labelsep}{0.5em}
} }

\newcommand{\squishend}{
  \end{list}
}

\setcopyright{acmlicensed}
\copyrightyear{2022}
\acmYear{2022}
\acmConference[CIKM '22]{Proceedings of the 31st ACM International Conference on Information and Knowledge Management}{October 17--21, 2022}{Hybrid Event, Atlanta, Georgia, USA}
\acmBooktitle{Proceedings of the 31st ACM International Conference on Information and Knowledge Management (CIKM '22), October 17--21, 2022, Hybrid Event, Atlanta, Georgia, USA}\acmDOI{10.1145/XXXXXXXXXX}
\acmISBN{XXXXXXX-XX/20/10}
\author{Marco Minici}
\email{marco.minici@icar.cnr.it}
\orcid{0000-0002-9641-8916}
\affiliation{%
  \institution{University of Pisa, Pisa, Italy}
  \institution{ICAR-CNR, Rende(CS), Italy}
}

\author{Federico Cinus}
\email{cinus@diag.uniroma1.it}
\orcid{0000-0002-6696-9637}
\affiliation{%
  \institution{Sapienza University, Rome, Italy}
  \institution{ISI Foundation, Turin, Italy}
}

\author{Corrado Monti}
\email{corrado.monti@centai.eu}
\orcid{1234-5678-9012}
\affiliation{%
 \institution{CENTAI, Turin, Italy}
}

\author{Francesco Bonchi}
\email{bonchi@centai.eu}
\orcid{1234-5678-9012}
\affiliation{%
  \institution{CENTAI, Turin, Italy}
  \institution{Eurecat, Barcelona, Spain}
}

\author{Giuseppe Manco}
\email{giuseppe.manco@icar.cnr.it}
\orcid{0000-0001-9672-3833}
\affiliation{%
  \institution{ICAR-CNR, Rende(CS), Italy}
}

\begin{document}

\title{Cascade-based Echo Chamber Detection}

\begin{abstract}
Despite echo chambers in social media have been under considerable scrutiny, general models for their detection and analysis are missing.
In this work, we aim to fill this gap by proposing a probabilistic generative model that explains social media footprints---i.e., social network structure and propagations of information---through a set of latent communities, characterized by a degree of echo-chamber behavior and by an opinion polarity.
Specifically, echo chambers are modeled as communities that are permeable to pieces of information with similar ideological polarity, and impermeable to information of opposed leaning: this allows discriminating echo chambers from communities that lack a clear ideological alignment.

To learn the model parameters we propose a scalable, stochastic adaptation of the Generalized Expectation Maximization algorithm, that optimizes the joint likelihood of observing social connections and information propagation. Experiments on synthetic data show that our algorithm is able to correctly reconstruct ground-truth latent communities with their degree of echo-chamber behavior and opinion polarity. Experiments on real-world data about polarized social and political debates, such as the Brexit referendum or the COVID-19 vaccine campaign, confirm the effectiveness of our proposal in detecting echo chambers.
Finally, we show how our model can improve accuracy in auxiliary predictive tasks, such as stance detection and prediction of future propagations.

\end{abstract}

\bigskip \bigskip \bigskip 


\bigskip \bigskip \bigskip 


\bigskip \bigskip \bigskip 

\maketitle \sloppy

\section{Introduction}
\label{sec:intro}

Social-media platforms have substantially altered the
landscape of societal debates. By delivering an extremely large
amount of content to online users, they enable quick and easy access to information and facilitate participation in public debates. This positive effect is intertwined by the growing phenomenon that online political discourses, especially on socially relevant issues, tend to fragment and polarize opinions. As a result, the propagation of information is affected by users' propensity to select and promote claims that adhere to their beliefs and ignore or even contrast dissenting information.

The \emph{``echo chamber''} effect in social media refers to groups of users that, by being exposed solely to like-minded individuals, tend to reinforce each other's pre-existing opinions.
This effect has been put under scrutiny as a possible culprit of increased polarization and radicalization~\cite{garrett2009echo}.
Thus, several studies have been devoted to providing empirical evidence of the existence of echo chambers~\cite{garimella2018quantifying,cinelli2021echo,de2021no}, with variable results, depending on the specific platforms and contexts. For instance, on Reddit echo chambers seem to be less prominent~\cite{de2021no}, while
on Twitter, it has been shown that information propagates in well-separated echo chambers~\cite{conover2011political,lai2019stance,cossard2020falling}.
However, this literature proposes ad hoc approaches for the detection of echo chambers in specific platforms and contexts, while a ground-up approach to detect echo chambers through a formal model of their behavior is still missing.

Prior studies~\cite{BarbieriBM13,MehmoodBBU13,barbieri2013cascade,B+16,barbieri2016efficient,R+18,ZHANG202286} have explored the role of communities in information propagation. The underlying assumption of these studies is that a user's activities and social connections are the visible effects of a latent stochastic diffusion process governed by community-level causal factors. As a result, the proposed models successfully devise communities through the lenses of social contagion and are able to characterize community membership in terms of propensity to filter and/or promote community-relevant information. Unfortunately, these models do not take into account the latent relationships between polarization and information diffusion that justify the formation of ideological groups and ultimately characterize the echo-chamber effect.

Inspired by this family of approaches for community detection, in this paper \emph{we tackle the problem of detecting echo chambers by observing the way polarized information propagates in a social network}. Similarly to communities, echo chambers are defined by groups of nodes who interact and exchange information in a social network. However, while communities in networks are simply defined as high-density clusters in the social graph, echo chambers only need enough structure to allow information to propagate, without the high density typical of a close-knit group of friends. Moreover, as put by \citet{alatawi2021survey}, an echo chamber can be defined as a community that spontaneously emerges as the most effective for spreading polarized content, where conflicting opinions are ignored or even discredited.
In other terms, we expect echo chambers to facilitate the flow, through its internals, of information that is ideologically aligned with its opinion, while preventing the flow of information with an opposed leaning.
Based on this intuition of echo-chamber behavior, we introduce a generative probabilistic model
that encodes, in probabilistic terms, the following core concepts: \emph{(i)} Each community is characterized by a measurable degree of polarization and highly polarized communities represent echo chambers. \emph{(ii)} Analogously, the participation of users to a specific community can be measured by means of a given engagement degree. \emph{(iii)} A polarized cascade can only occur within an echo chamber, provided that the corresponding polarities are aligned. Furthermore, the likelihood of a user contributing to a cascade depends on their level of engagement in the corresponding community. \emph{(iv)} Social connections are likely to occur between community members;  however, it is possible to explain such connections differently according to whether the underlying latent community is considered an echo chamber or not.

Notably, the underlying latent parameters for such a model can be efficiently inferred by resorting to a suitable adaptation of the Generalized Expectation Maximization algorithm~\cite{bishop:2006:PRML}, which exploits samples of observable propagations and social connections to learn such parameters through an alternating gradient-based optimization strategy. As a result, the learning is scalable and likely to produce communities that can be amenable to coherent explanations in terms of echo-chamber behavior and opinion polarity.

\spara{Paper contributions and roadmap.}
Our technical contributions can be summarized as follows:
\squishlist
\item We propose a community-aware information propagation model that explains the creation of social links and the diffusion of items in terms of homogeneity and alignment within a polarized ideological space (§3).

\item We devise a scalable gradient-based optimization procedure
to learn both the communities and their degree of polarization, maximizing an approximation of the likelihood of a set of information cascades (§4).

\item We provide an extensive empirical evaluation of our proposal
on synthetic and real-world datasets and show that our inference algorithm is effective, provides meaningful and interpretable communities, and can be used to predict auxiliary tasks such as activation of users on a given cascade, or their stance (§5).
\squishend

 In the next section, we discuss relevant related work.

\section{Related work}
\label{sec:related}

\noindent \textbf{Echo chambers: causes and traits.}
The mechanisms behind the formation of echo chambers are still subject of investigation, however, three main phenomena seem to play key roles in the formation process: algorithmic recommendation (both content~\cite{fabbri2022rewiring}, and people~\cite{cinus2021effect} recommenders), confirmation bias~\cite{del2017modeling,quattrociocchi2016echo}, and homophily~\cite{jiang2021mechanisms}.
The literature also presents several characterizations of echo chambers highlighting their distinctive traits, such as distorted information patterns~\cite{jamieson2008echo}, users similarity growth~\cite{cinelli2021echo}, users psychological profiles~\cite{bessi2016personality}, the presence of polarization effects~\cite{garimella2018quantifying}, and the capacity of spreading polarized content~\cite{alatawi2021survey}.

\noindent \textbf{Echo chambers: detection.} Some prior works have analysed social media information to detect echo chambers in a variety of different platforms, without proposing a general method~\cite{cinelli2021echo,villa2021echo,quattrociocchi2016echo}.
Usually, a combination of different sources of information is considered, such as textual features (e.g., tweets and hashtags) and interaction network (e.g., retweets, mentions, follow).
While some studies (e.g.~\cite{calderon2019content}) focus mainly on the former, in this work we exploit, together with the social graph, the polarity of information pieces and their cascades, which have been shown to be more effective in detecting communities~\cite{prokhorenkova2022less} and user-level stances~\cite{aldayel2019your}.
Other researchers have approached the task of detecting echo chambers as a community detection task, thus exploiting the network structure, followed by an interpretation step to properly identify communities that have echo-chamber traits~\cite{guarino2019beyond,del2017mapping,cossard2020falling,cota2019quantifying,du2016echo}. In our model, the community structure and the level of echo chamberness of each community are learnt \emph{jointly}.
A recent effort by \citet{morini2021toward} adopts a hybrid approach, employing network-based methods together with NLP tools for content analysis.
In particular, the authors propose to first infer users’ ideology on a controversial issue, thus constructing a debate network, then detect polarity-homogeneous communities.
Their framework is the most similar to ours in terms of input and output:
nonetheless, it differs from our proposal as it does not exploit the mutual influence of social bonds
and cascades. Moreover, it lacks an explanation for the inferred echo chambers, as it is not based on a model, instead it is a concatenation of pre-existing techniques.

\noindent \textbf{Community detection from cascades.}
\citet{barbieri2013cascade} introduce a stochastic generative model that relies on a mixture of memberships to learn communities from observed cascades.
Similarly to our approach, learning is performed by means of Expectation-Maximization: the derived model can both output the overlapping communities and the users' level of participation.
Nevertheless, this approach does not take in consideration the polarity of content and the differences between normal social communities and echo chambers.
In the context of dynamic networks, \citet{he2021discovering} use cascade diffusion models  to discover overlapping communities.
In the same context, \citet{sattari2018cascade} propose a hybrid approach that relies on label propagation and cascades models to learn overlapping communities in dynamic networks.
A related research line has attempted inferring the communities from the information cascades \emph{only}, i.e., when the underlying network is not observable~\cite{barbieri2016efficient,prokhorenkova2022less}. Despite this major difference, the proposed approaches share important features with our work, such as inferring the individual membership by maximum-likelihood.
Finally, \citet{monti2021learning} have recently proposed to exploit cascades models to learn users' opinions/leanings in social networks. In particular, \cite{monti2021learning} introduces a stochastic model-based approach to learning, by gradient-based optimization, the ideological leaning of each user in a multidimensional ideological space. While their framework targets directly the user-level representation,
in our work we focus on the mesoscale community structure.
\begin{figure*}[]
    \centering
    \begin{subfigure}[t]{.3\linewidth}
      \centering\includegraphics[width=\linewidth]{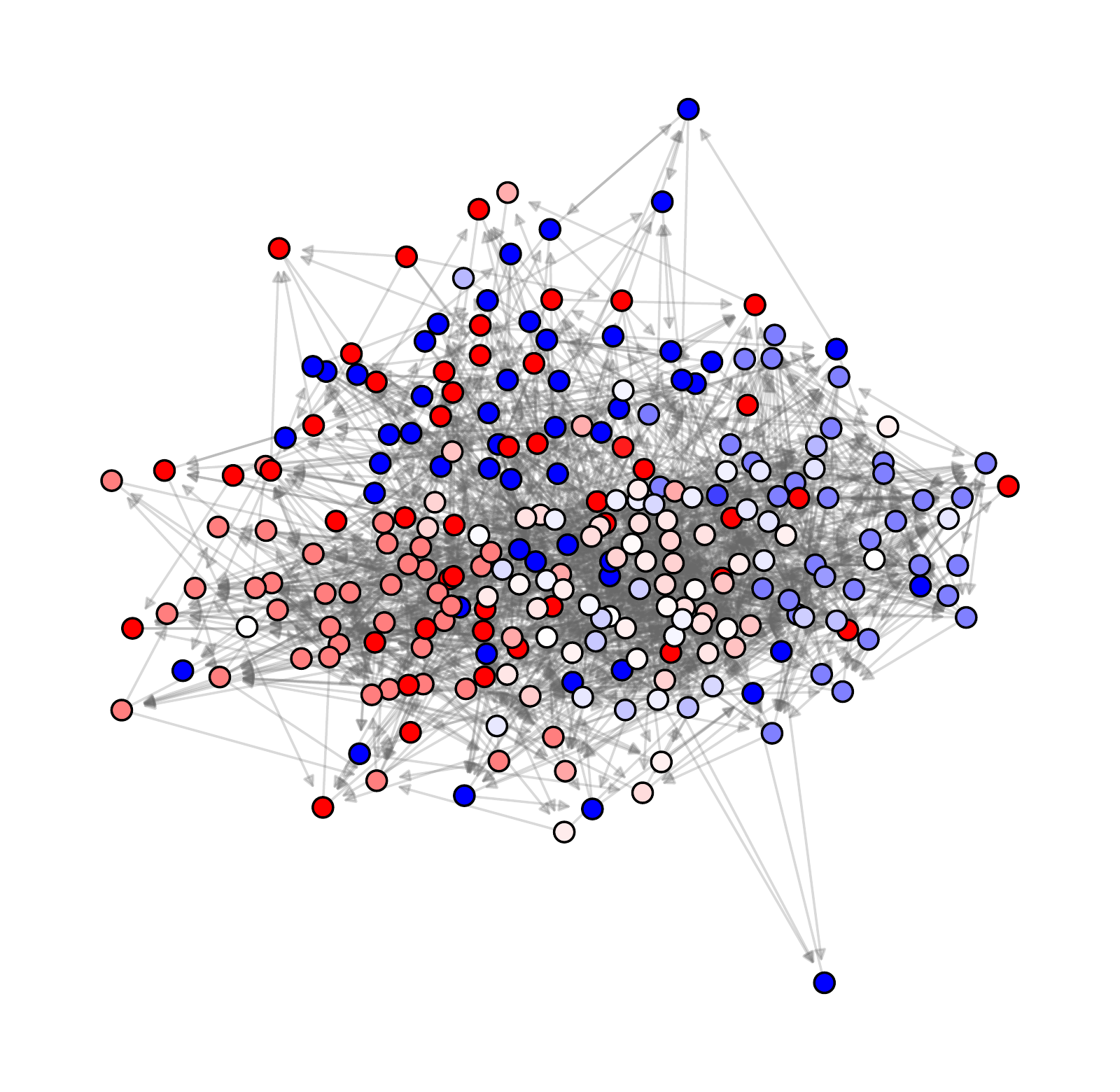}
      \vspace{-10mm}
      \caption{Social}
      \label{fig:examples-social}
    \end{subfigure}
    \begin{subfigure}[t]{.3\linewidth}
      \centering\includegraphics[width=\linewidth]{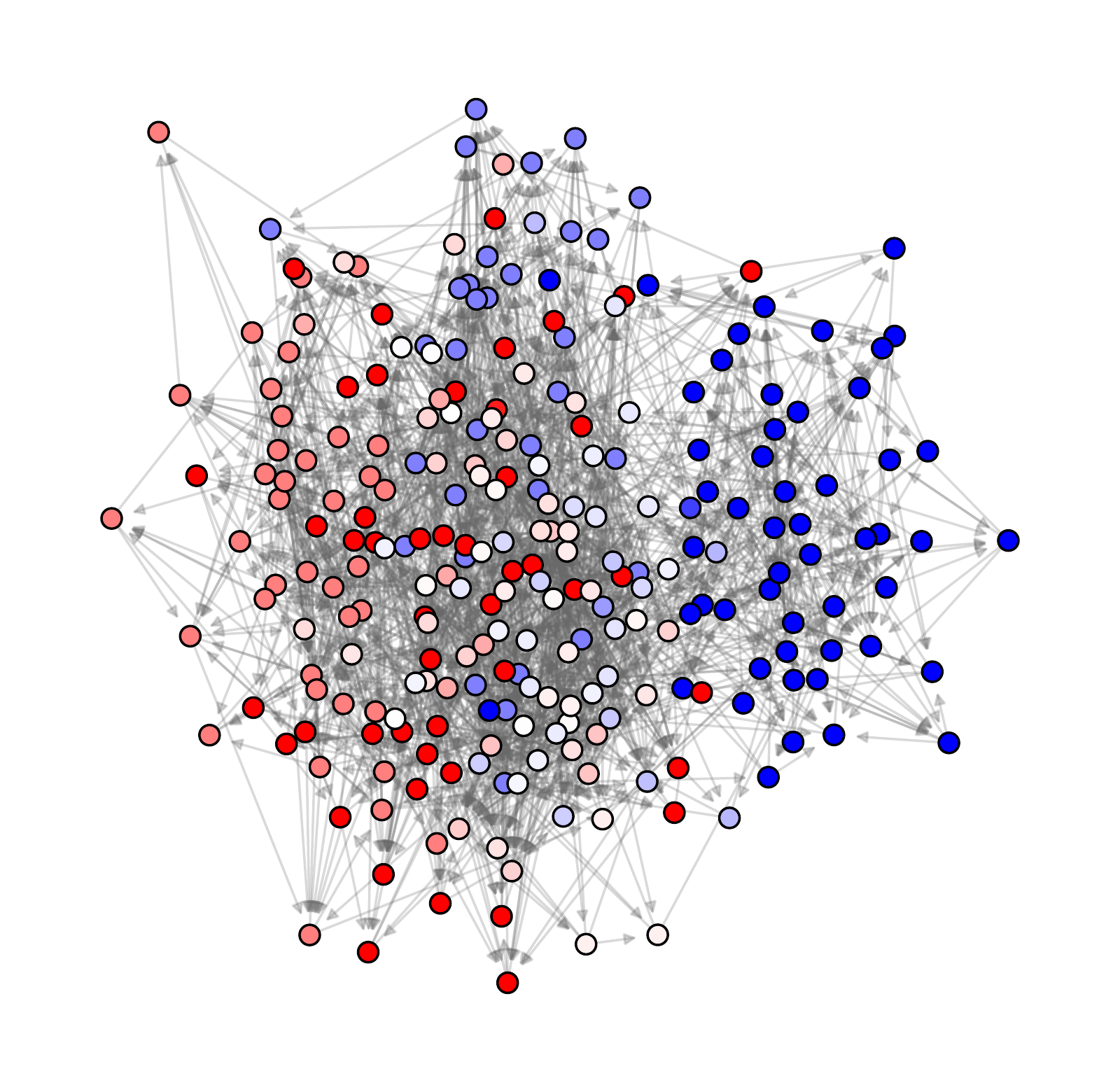}
     \vspace{-10mm}
      \caption{Balanced}
      \label{fig:examples-balanced}
    \end{subfigure}
    \begin{subfigure}[t]{.3\linewidth}
      \centering\includegraphics[width=\linewidth]{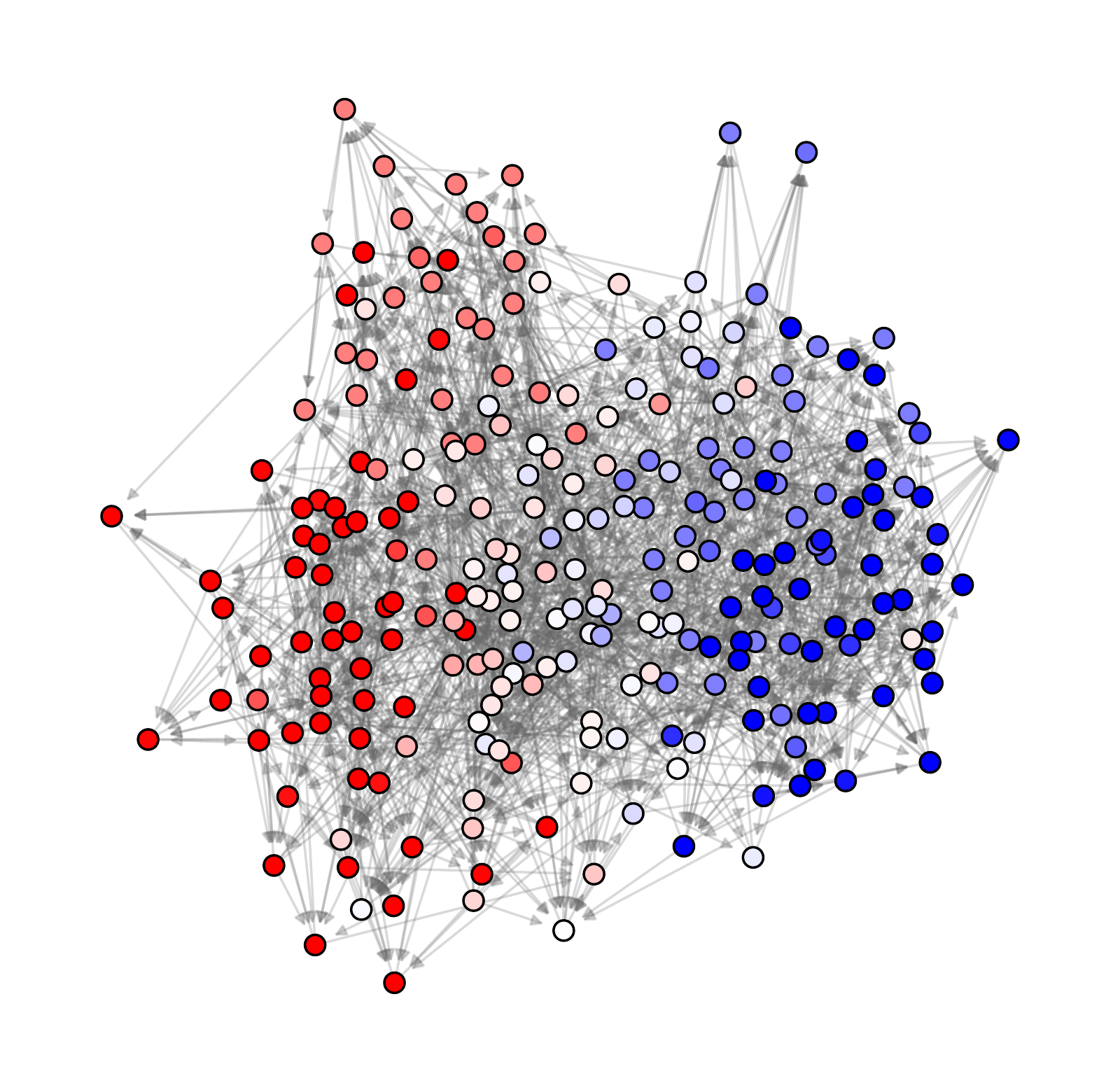}
      \vspace{-10mm}
      \caption{Polarized}
      \label{fig:examples-polarized}
    \end{subfigure}
    \vspace{\captionfigdist}
    \caption{Generated examples for the model with different values for $s, h$ in order to have different balances of echo chambers and social communities in the network: Social with $(s=16, h=8)$, Balanced with $(s=8, h=8)$, and Polarized with $(s=8, h=16)$.}
    \label{fig:examples-polarization}
\vspace{\figtextdist}
\end{figure*} 

\section{Modeling Echo Chambers}
\label{sec:model}
We next introduce our probabilistic generative model of echo chambers, which has two main goals.
First, to provide a generative description of the echo-chamber phenomenon.
Second, to draw on this description in order to identify echo chambers in the wild.
We start by identifying some variables as observables ---representing the evidence of the phenomena we aim at modeling --- and others as latent--- i.e., the components that explain the observables.

\spara{Observables.}
We consider the following input:
\begin{enumerate}
  \item A directed social graph $G= (V,E)$, where $V$ represents a set of social media users, and $E$ a set of links, where $(u,v) \in E$ represents the fact that user $u$ is followed by user $v$.
  \item A set of items \I, where each item $i \in \I$ is labeled with a polarity $p_i \in [-1,1]$, that characterize its ideological content (e.g., with respect to a given political axis). As we are interested in modeling echo chambers, we assume that our input set $\I$ only contains polarized items, i.e., whose $p_i$ value is not close to 0.
  \item For each item $i \in \I$, its cascade $\D_i \subseteq V$ in the social graph, i.e.,  the set of nodes that propagated (or consumed) item $i$.
\end{enumerate}

\spara{Latent variables.}
We assume that nodes in $V$ can be grouped in latent communities.
Some of these communities are \emph{echo chambers}; that is, they facilitate the flow, through its internals, of information that is ideologically aligned with its opinion, while preventing the flow of information with an opposed leaning.
By contrast, we call \emph{social communities} those communities that are likely to incorporate ideologically heterogeneous nodes.

Given a set $C$ of latent communities, the value $\eta_c\in [-1, 1]$ (with $c\in \{1, \ldots, K\}$)  indicates
both the polarization and the degree of echo-chamber behavior of the community $c$.
In particular, the value $|\eta_c| = 1$ indicates an ideal echo chamber community, while $|\eta_c| = 0$ indicate the perfect social community.

We assume that each observation (i.e., the links of the social graph and the cascades) is the result of a stochastic process where people act in the network according to their fuzzy membership to latent communities. More specifically, we assume two prior components for a given node $u\in V$ and community $c$:
\begin{itemize}
  \item $\theta_{c,u}\in [0,1]$ represents the level of \emph{polarized engagement} of user $u$ in a echo chamber $c$;
  \item $\phi_{c,u}\in [0,1]$ is the level of \emph{social engagement} of user $u$ in a social community $c$.
\end{itemize}
Both $\theta_{c}$ and $\phi_{c}$ represents categorical distributions and model how likely is that a user $u$ contributes to $c$. %
Each phenomenon in the social network can be explained by the above latents: a link $(u, v)$ can only be observed if $u$ and $v$ are part of the same close-knit community or if they are part of echo chambers with the same polarity.
Similarly, each item $i$ is produced by a community $c$ and it propagates by flowing through the nodes of that community.

\subsection{Modeling links and propagations}
We use the above latent variables to devise a stochastic process generating the observables.
As discussed previously, we assume that echo chambers facilitate the flow of information
which is ideologically aligned with their opinion, while preventing the
flow of information with an opposed leaning.
Following this assumption, the propagation of an item $i$ is generated by considering its polarity $p_i$ and the polarity of the echo chamber $\eta_c$.
Specifically, echo chambers only allow items with the same polarity; therefore, a propagation is allowed only if $\operatorname{sign}(p_i) = \operatorname{sign}(\eta_c)$, and is allowed with a probability depending on the degree of echo-chamber behavior of $c$; i.e., a Bernoulli trial with probability $|\eta_c|$.
Finally, it depends on how strong is the item polarity, i.e. a Bernoulli trial with probability $|p_i|$.
If these conditions are respected, a node is chosen from the categorical distribution of the community, i.e. $u \sim \operatorname{Cat}(\theta_c)$.
This process gives rise to the following likelihood of observing a propagation $\D_i$ with polarity~$p_i$:
\begin{equation}
  P(\D_i | c) =  \max(0, p_i \cdot \eta_c) \prod_{u\in \D_i} \theta_{c,u}. %
  \label{eq:d-given-c}
\end{equation}
In other words, item propagations can only be explained by an alignment between the item polarity and the sign of $\eta_c$.
In fact, the term $\max(0, p_i \cdot \eta_c)$ is only positive when both $p_i$ and $\eta_c$ exhibit the same sign.
If this is the case, users can contribute to the propagation according to their degree of echo-chamber involvement.

Similarly, social links in $G$ are generated as follows.
Each community $c$ chooses whether it is an echo chamber with a Bernoulli random trial with probability $|\eta_c|$.
If it is, it will create a link $(u, v) \in E$ by extracting two nodes $u, v$ by using the $\theta_c$, i.e. $u \sim \operatorname{Cat}(\theta_c)$.
Otherwise, it will do so by using the social engagement $\phi_c$.
This procedure defines the probability of each link $(u, v)$ given that it was latently generated by a given community $c$ as
\begin{equation}
  P \big((u,v) \in E| c \big) = |\eta_c|\cdot \theta_{c,u} \theta_{c,v} + (1 - |\eta_c|)\cdot \phi_{c,u} \phi_{c,v}.
  \label{eq:l-given-c}
\end{equation}
The above probability follows from this stochastic process where users contribute to the underlying community according to the polarity of the community itself.
In fact, $|\eta_c|$ represents a characterization of the community either as an echo chamber (and consequently links forms with $\theta_c$ and propagations are possible) or as a social community (links occur thanks to $\phi_c$).

\spara{Latent priors.}
Equations~\ref{eq:d-given-c} and \ref{eq:l-given-c} model conditional probabilities for links and propagations, given $c$.
In order to specify the unconditional likelihood, we introduce the categorical priors $\pi_{\ell}$ and $\pi_f$; those define the probability of creating respectively links and propagations.
The term $\pi_{\ell}(c)$ (resp. $\pi_f(c)$) represents the prior probability of a link (resp. a propagation) within $c$.
However, according to our assumption, the probability $\pi_f(c)$ strongly depends on the polarity $\eta_c$: when $\eta_c \approx 0$, propagations cannot be explained through $c$.
This constraint can be enforced by introducing the Dirichlet priors $\alpha^f$ and $\alpha^l$ defined as
\begin{equation}\label{eq:pi-priors}
\alpha^f_c = h \cdot |\eta_c| + \epsilon, \qquad\qquad \alpha^l_c = s\cdot (1-|\eta_c|) + h\cdot |\eta_c|
\end{equation}
where hyperparameters $s > 0$, $h > 0$ represent the prior importance of social and echo chamber communities (respectively) in generating links, and $\epsilon$ is a regularization value (e.g. $10^{-5}$).
Therefore, we can generate $\pi_f$ and $\pi_\ell$ through sampling from Dirichlet distributions parameterized by $\alpha^f$ and $\alpha^l$.

\spara{Likelihood.} We can finally specify the likelihood for both links and propagations. Given the model parameters $\Theta = \{\theta, \phi,\eta\}$ and the hyperparameters $s$ and $h$, we have:
\begin{equation}\label{eq:llk}
\begin{split}
    P(\ell|\Theta; s, h) & =  \int \left\{\sum_c P(\ell|c) \pi_\ell(c) \right\}\mathit{Dir}\left(\pi_\ell; \alpha^l\right) \mathrm d\; \pi_\ell \\
    P(\D_i|\Theta; s, h) & =  \int \left\{\sum_c P(\D_i|c) \pi_f(c) \right\}\mathit{Dir}\left(\pi_f; \alpha^p\right) \mathrm d\; \pi_f
\end{split}
\end{equation}
For readers' convenience, we provide a notation reference in \Cref{tab:notation}.

\begin{table}[b]
    \caption{Notation reference. %
    \label{tab:notation} }
    \vspace{\captionfigdist}
    \small
    \begin{adjustbox}{max width=\textwidth,center}
    \begin{tabular}{ll}
        \toprule
        Variable & Meaning \\
        \midrule
         $\eta_c$ & Polarity of community $c$ \\
         $\theta_{c,u}$ & Polarized engagement of user $u$ in community $c$ \\
         $\phi_{c,u}$ & Social engagement of user $u$ in community $c$\\
         $p_i$ & Polarity of item $i$ \\
         $\pi_\ell(c)$ & Prior link probability in community $c$\\
         $\pi_f(c)$ & Prior propagation probability in community $c$ \\
         $\alpha^l_c$ & Parameter of the Dirichlet distrib. that defines $\pi_\ell$ \\
         $\alpha^f_c$ & Parameter of the Dirichlet distrib. that defines $\pi_f$ \\
         $h$ & Link generation strength of echo chambers \\
         $s$ & Link generation strength of social communities \\
         $\gamma_{\ell, c}$ & Posterior to observe a link $l$ in community $c$ \\
         $\xi_{\p,c}$ & Posterior to observe a propagation $\p$ in community $c$ \\
        \bottomrule
    \end{tabular}
    \end{adjustbox}
\vspace{\figtextdist}
\end{table}

\subsection{Generative process}
\label{sec:generative-process}
\begin{figure}
\vspace{-3mm}
  \begin{subfigure}[t]{.49\linewidth}
    \centering%
    \includegraphics[trim={0 14mm 9mm 0},clip,width=\linewidth]{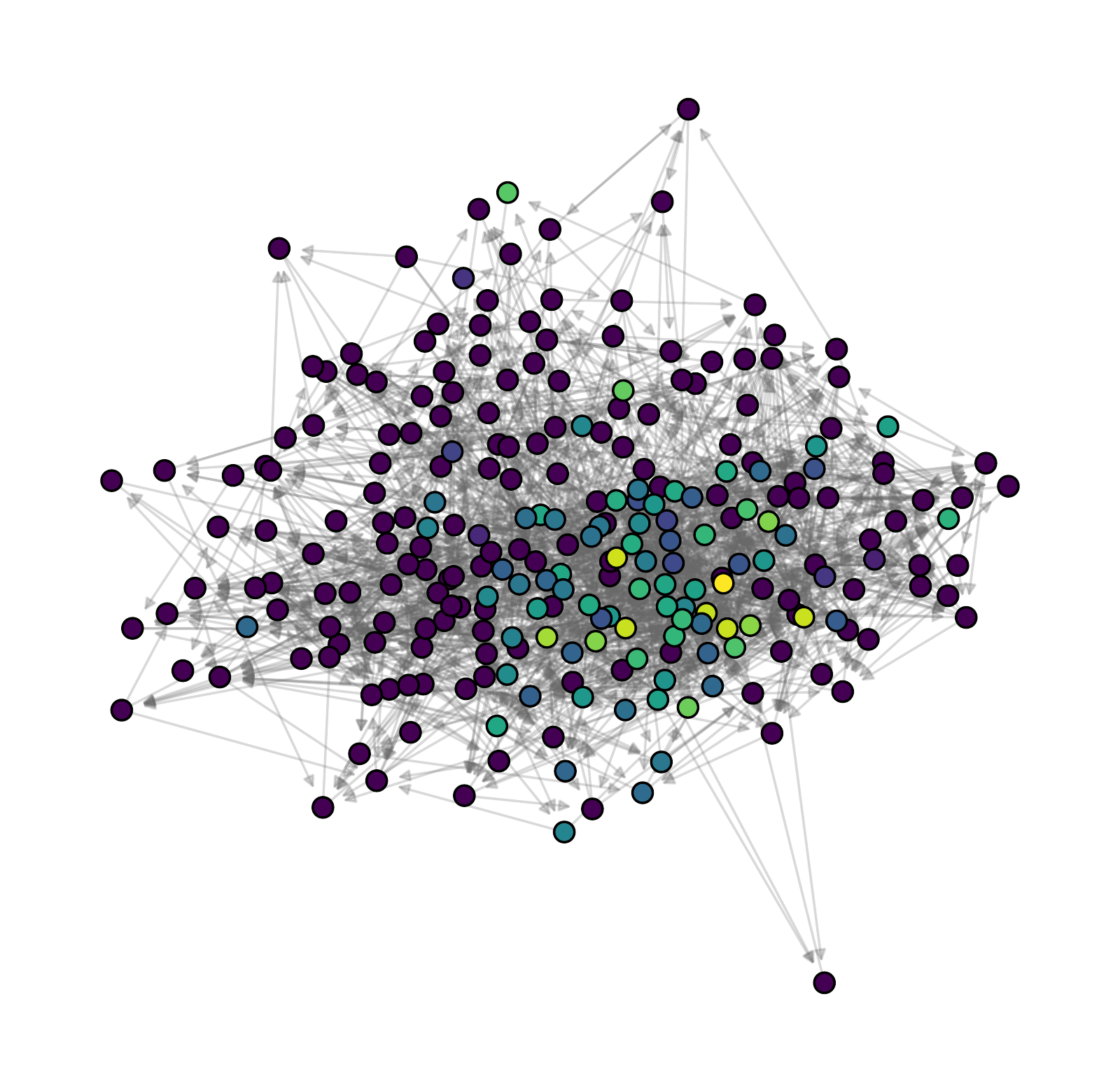}
    \caption{Community}
    \label{fig:examples-community}
  \end{subfigure}
  \begin{subfigure}[t]{.49\linewidth}
    \centering%
    \includegraphics[trim={9mm 14mm 0 0},clip,width=\linewidth]{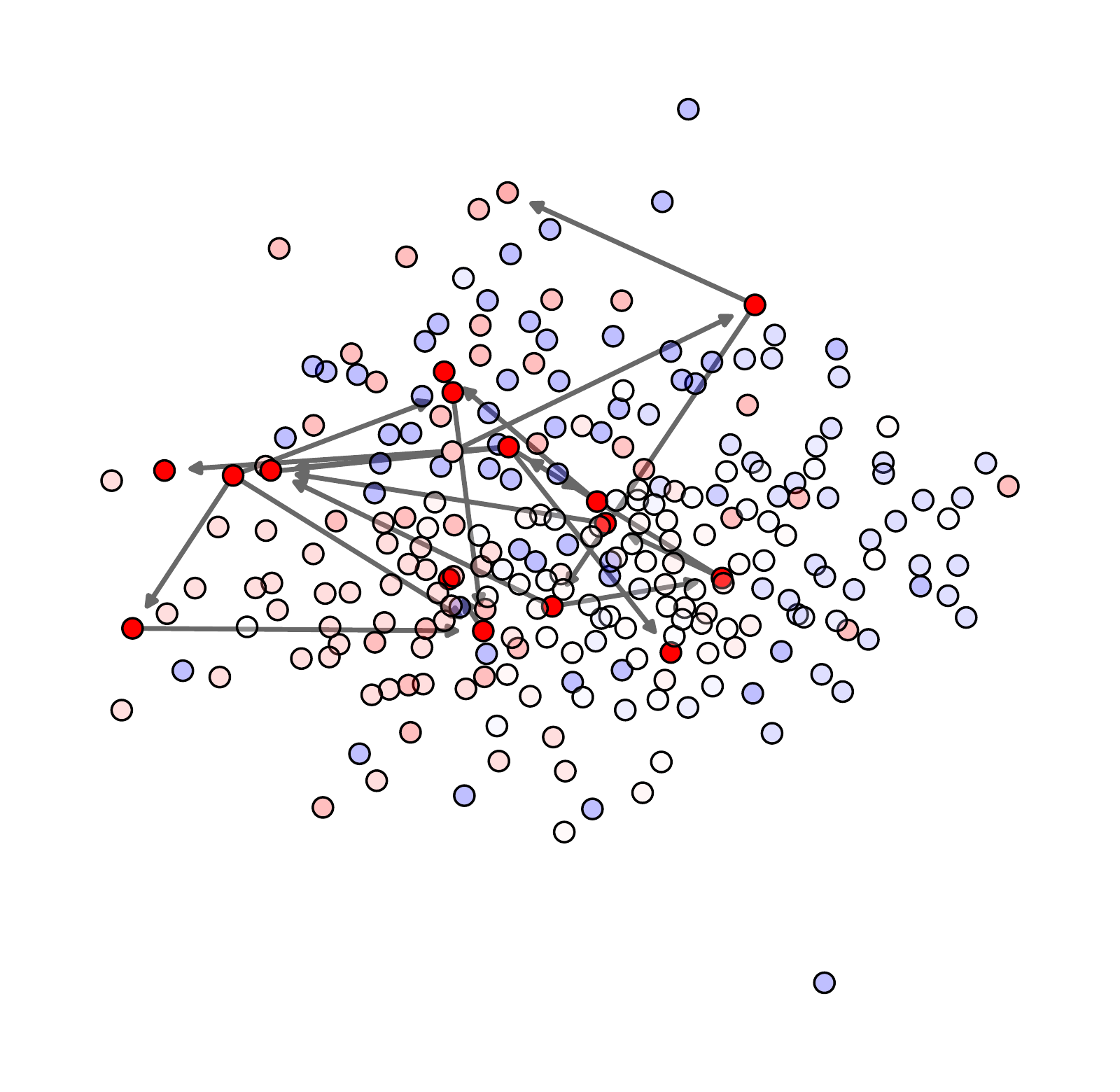}
    \caption{Propagation}
    \label{fig:examples-propagation}
  \end{subfigure}
  \vspace{\captionfigdist}
  \caption{Example community and propagation generated for the network %
  represented in \Cref{fig:examples-social}. Left (\subref{fig:examples-community}) shows the membership for one social-type community with shades of green. Right (\subref{fig:examples-propagation}) shows one propagation, spreading within an echo-chamber community with homogeneous polarity, where polarity is represented by node colors.}
  \label{fig:examples-comm}
   \vspace{-3mm}
\end{figure}

We can summarize the aforementioned procedure as a simple generative stochastic process for data generation that adheres to the aforementioned modeling assumptions.
The process assumes that $V$ and $I$ are given; then, based on the model parameters it generates both links and item propagations with the following processes.

\spara{Links.} The generative process for a link $\ell$ is:

\begin{enumerate}[(i)]
  \item Pick a community $c_\ell \sim \operatorname{Cat}(\pi_{\ell})$.
  \item Pick $y_c \sim \mathit{Bernoulli}(|\eta_{c_\ell}|)$ (whether $c_\ell$ is an echo chamber).
  \item If $y_c>0$,  Pick two nodes $u, v \sim \operatorname{Cat}(\theta_c)$.
  \item Else, pick two nodes $u, v \sim \operatorname{Cat}(\phi_c)$.
  \item Add the arc $(u, v)$ to $E$.
\end{enumerate}

\spara{Items.}
The generative process for the propagation of an item $i$ with polarity $p_i$ is:

\begin{enumerate}[(i)]
    \item Repeat:
    \begin{itemize}
        \item pick $c \sim \operatorname{Cat}(\pi_f)$;
        \item $y_i \sim \mathit{Bernoulli}(g_i)$ where $
    g_i(c) = \max(0, p_i \cdot \eta_c);
  $
    \end{itemize}
    until $y_i >0$.
    \item Pick a user
    $u \sim \operatorname{Cat} \left(\theta_{c}  \right)$ and let $\mathbb{D}_i = \{u\}$.
    \item Repeat:
    \begin{itemize}
    \item \label{item:repeat-from} let $F_{i}= \{u | (v,u) \in E, v \in D_i, u \not\in D_i\}$;
    \item pick the next user
    $$u \sim \operatorname{Cat} \left(
    \theta_{c} \cdot [u \in F_{i}] \right);$$
    \item add $u$ to the set of activated nodes $D_i$;
  \end{itemize}
  until $F_{i} = \emptyset$ or $D_i$ has reached a given size.
\end{enumerate}

\subsection{Generated networks}
\label{sec:generated-networks}
Here, we show and analyze some networks generated by the generative process we devised.
We consider three different sets of parameters $s, h$ (defined in Equation~\ref{eq:pi-priors}), in order to obtain a varying degree of echo-chamber behavior in the network.
In all these networks, we generate 5 communities with a fixed $\eta = [ -1, -0.5, 0.0, 0.5, 1]$: two opposing echo chambers, a purely social community, and two cases in-between.
Then, we randomly generate $\theta$ and $\phi$ for $N=256$ nodes.
Considering $s \in \{-1, 1\}$ we define two echo-chamber priors
$\alpha^s_{c} = \max(0, s \cdot \eta_c) \cdot \sigma_s + \epsilon$ where $\sigma_s$ are concentration parameters.
Analogously, we define a social-type prior for $\phi$ with parameters $\alpha^{0} = (1 - |\eta|)\cdot \sigma_0$.
Then, for each node $u$ we obtain the membership
$\tilde{\theta}_u = U_p \cdot P + (1-U_p)\cdot U_n \cdot N$ and
$\tilde{\phi}_u = (1-U_p\cdot U_n)\cdot S$,
where $P \sim \operatorname{Dir}(\alpha^p)$,
$N \sim \operatorname{Dir}(\alpha^n)$,
$S\sim\operatorname{Dir}(\alpha^0)$ and
$U_p, U_n \sim \operatorname{Bernoulli}(\delta)$
(we set $\delta = .3$ in experiments).
Finally, we generate the network according to our generative procedure, producing 2048 links (an average of $8$ links per node). Similarly, we generate 2048 propagations of items; to generate item polarities, we draw them as $p_i = 2 X - 1$ with $X \sim \text{Beta}(\mu, \mu)$, where $\mu=0.25$ is a parameter regulating the ideological strength of the generated items.
Figure~\ref{fig:examples-polarization} shows the three graphs obtained with these different settings of $(s, h)$.
In each visualization, the color gradient represents the polarities for each node, obtained as $\eta \cdot\theta$ (i.e., the weighted average polarity of each community for a given node).
We observe that the first graph (\Cref{fig:examples-social}) appears not to be shaped by polarized communities, but instead nodes with similar polarities are scattered across the network.
In Figure~\ref{fig:examples-comm} we further explore the first generated network from \Cref{fig:examples-social}, where links are predominantly generated by social-type communities.
We depict two features of this data set: a social community, and a propagation.
We see that social communities, while embedded in a dense network (as in real-world data sets), are closely-knitted in the network.
Nevertheless, propagations still happen inside echo-chamber communities, spreading across ideologically aligned nodes.

\section{Model Learning}
\label{sec:algorithms}

Given $G$ and $\I$, the optimal $\Theta$ parameters can be learned by maximizing the total likelihood
$$
P(E, \I |\Theta) = \prod_{\ell \in E} P(\ell|\Theta) \prod_{\substack{\D_i\\i \in \I}} P(\D_i|\Theta).
$$
First of all, we notice that Equation~\ref{eq:llk} can be simplified by exploiting the conjugacy of the Dirichlet Distribution~\cite{bishop:2006:PRML}:
\begin{equation*}%
        P(\ell|\Theta)  =  \sum_c P(\ell|c) \pi_\ell(c) , \ \ \
     P(\D_i|\Theta)  =  \sum_c P(\D_i|c)\pi_f(c) ,
\end{equation*}
where
\begin{equation}\label{eq:priors}
     \pi_\ell(c) = \frac{\alpha^l_c}{\sum_{c'} \alpha^l_{c'}},\ \ \ \pi_f(c) = \frac{\alpha^f_c}{\sum_{c'} \alpha^f_{c'}}.
\end{equation}
A potential problem with the resulting optimization problem is represented by the contribution of each propagation in the total likelihood.

By comparing Equations~\ref{eq:d-given-c} and~\ref{eq:l-given-c}, we observe that the probability of a link embeds a product over two probabilities, whereas by contrast the probability of a propagation embeds the product over multiple probabilities.
Thus, long propagations have very low probability and as a consequence the whole learning process is dominated by link probabilities.
This issue makes it difficult to effectively learn the latent variable $\theta$, and consequently the detection of echo chambers.

This problem can be addressed by resorting to a surrogate version of the above likelihood.
In practice, we can consider the weighted multi-graph $G^{\I}=(V, E^{\I})$ induced by all propagations, with $\p(u,v,p_i)\in E^{\I}$ if $u,v\in \D_i$ representing a \emph{sharing link} (i.e., both $u$ and $v$ share an item $i$, characterized by polarity $p_i$).
Then, the probability of observing such a link can be directly adapted from Equation~\ref{eq:d-given-c}:
\begin{equation*}\label{eq:co-occurrence-given-c}
P(\p(u,v,p)|c) = \max(0, p \cdot \eta_c) \theta_{c,u}\theta_{c,v}
\end{equation*}

Thus, the total likelihood can be rewritten into $P(E, E^{\I}) = \prod_{\ell \in E} P(\ell|\Theta) \prod_{\p \in E^{\I}} P(\p|\Theta)$ that allows a more balanced approach through stochastic backpropagation, where each batch can include a sample of both social connections and sharing links.
We further simplify the optimization problem by resorting to a variational approximation.
Let $X\subseteq E\cup E^\I$ be a batch of social connections and sharing links,  $Y$ be a set of corresponding binary variables representing the latent community assignment for both social connections and sharing links;
that is, $y_{\ell,c}=1$ (resp. $y_{\p,c}=1$) if $\ell$ (resp. $\p$) is associated to community $c$.
Observe that
\begin{align*}
\log P(X, Y|\Theta) =  \sum_{\ell, \p \in X} \sum_c \Big\{ & \log P(\ell|\Theta, c) + \log \pi_\ell(c) \\
& + \log P(\p|\Theta, c) + \log \pi_f(c)\Big\},
\end{align*}
and define
\begin{equation}\label{eq:complete-llk}
\begin{split}
\mathcal{Q}(\Theta, & \Theta'|X)  =   \mathbb{E}_{Y|X, \Theta'}\left[\log P(X, Y|\Theta)\right] \\
= & \sum_{\ell,\p\in X} \sum_c \Big\{P(y_{\ell,c}|\ell, \Theta')\left(\log P(\ell|\Theta, c) + \log \pi_\ell(c)\right) \\
& \qquad\qquad +  P(y_{\p,c}|\p, \Theta')\left(\log P(\p|\Theta, c) + \log \pi_f(c)\right)\Big\}.
\end{split}
\end{equation}
Notably, whenever  $\mathcal{Q}(\Theta, \Theta'|X) \geq \mathcal{Q}(\Theta', \Theta'|X)$, then $\log P(X|\Theta) \geq \log P(X|\Theta')$.
In fact,
\begin{align*}
    \log P(X|\Theta) = & \mathbb{E}_{Y|X, \Theta'}\left[\log P(X|\Theta)\right] \\
    = & \mathbb{E}_{Y|X, \Theta'}\left[\log \frac{P(X, Y|\Theta)}{P(Y|X,\Theta)}\right] \\
    = & \mathcal{Q}(\Theta, \Theta'|X) - \mathbb{E}_{Y|X, \Theta'}\left[\log P(Y|X,\Theta)\right] \\
    \geq & \mathcal{Q}(\Theta', \Theta'|X) - \mathbb{E}_{Y|X, \Theta'}\left[\log P(Y|X,\Theta)\right] && \mbox{(a)}\\
    = & \mathcal{Q}(\Theta', \Theta'|X) - \mathbb{E}_{Y|X, \Theta'}\left[\log P(Y|X,\Theta')\right] \\
    & - \mathbb{E}_{Y|X, \Theta'}\left[\log \frac{P(Y|X, \Theta)}{P(Y|X,\Theta')}\right] \\
    \geq & \mathcal{Q}(\Theta', \Theta'|X) - \mathbb{E}_{Y|X, \Theta'}\left[\log P(Y|X,\Theta')\right] && \mbox{(b)}\\
    = & \mathbb{E}_{Y|X, \Theta'}\left[\log \frac{P(X, Y|\Theta')}{P(Y|X,\Theta')}\right] \\
    = & \log P(X|\Theta'),
\end{align*}
where (a) holds by hypothesis, and (b) by Jensen's inequality.
This enables an iterative optimization strategy where, for each iteration $t$, we sample a batch $X$ of social connections and sharing links, and then apply the following alternating steps:
\begin{itemize}
    \item (\emph{Expectation}) For each $\ell, \p\in X$ and community $c$, compute the posteriors
    \begin{equation}\label{eq:posteriors}
        \begin{split}
            \gamma_{\ell, c} \equiv P(y_{\ell,c}|\ell, \Theta^{(t)}) = \frac{P(\ell|c)\pi_\ell(c)}{\sum_{\hat{c}} P(\ell|\hat{c})\pi_\ell(\hat{c})}\\
            \xi_{\p,c} \equiv P(y_{\p,c}|\p, \Theta^{(t)})= \frac{P(\p|c)\pi_f(c)}{\sum_{\hat{c}} P(\p|\hat{c})\pi_f(\hat{c})},
        \end{split}
    \end{equation} given the current parameter set $\Theta^{(t)} = \{\eta^{(t)},\theta^{(t)} \phi^{(t)}\}$
    \item (\emph{Optimization}) Ascend the gradient $\nabla_{\Theta} \mathcal{Q}(\Theta, \Theta^{(t)}|X)$ to obtain $\Theta^{(t+1)}$.
\end{itemize}

The whole procedure, dubbed ECD (Echo Chamber Detection), is described in~\Cref{alg:algorithm}.
\begin{algorithm}[t]
   \footnotesize
    \caption{ECD Inference}
    \label{alg:algorithm}
    \begin{flushleft}
    \algorithmicrequire \; Graph $G = (V, E)$; Sharing links $E\^I$. \\
    \textbf{Hyper-parameters}: number of communities $C$, social prior size $s$, \\
    \hspace{2.1cm} echo-chamber prior size $h$, learning rate $\lambda$, \\
    \hspace{2.1cm} number of optimization steps for each iteration $H$. \\
    \algorithmicensure \;polarities $\eta$, memberships $\theta$ and $\phi$.
    \begin{algorithmic}[1]
            \State Randomly initialize $\Theta^{(0)} = \{\eta^{(0)}, \theta^{(0)}, \phi^{(0)}\}$ and set $t=0$.
            \Repeat
            \State let $\Theta^{(*)} = \Theta^{(t)}$
            \For{$w\in \{1, \ldots, H\}$}
                \State Sample $X$ from $E\cup E^\I$.
                \ForEach{$\ell, \p \in X$ and $c\in \{1, \ldots, C\}$}
                    \State Compute posteriors $\gamma_{\ell,c}$ and $\xi_{\p, c}$ according according to Eqs.~\ref{eq:posteriors},~\ref{eq:priors} and the current parameters $\Theta^{(t)}$.
                    \Comment{E Step}
                \EndFor
                \State Compute the expected likelihood $\mathcal{Q}$ according to Eqs.~\ref{eq:complete-llk} and~\ref{eq:priors} and the posteriors $\gamma$ and $\xi$.
                \State Update the parameters:
                \Comment{M Step}
                $$
                \Theta^{(*)} = \Theta^{(*)} + \lambda \nabla_{\Theta} \mathcal{Q}(\Theta^{(*)}, \Theta^{(t)}|X)
                $$
            \EndFor
            \State Set $\Theta^{(t+1)} = \Theta^{(*)}$ and increase $t$:
        \Until{convergence}
    \end{algorithmic}
  \end{flushleft}
  \normalsize
\vspace{\figtextdist}
\end{algorithm}

It converges to a local minimum, for a sufficiently small learning rate, since it preserves the general properties of stochastic backpropagation.
In fact, although there is no guarantee that the improvement in the likelihood of the current batch corresponds to an improvement in the likelihood of the whole set of observables, this property occurs on average and can eventually be improved by, at each iteration, freezing $\Theta^{(t)}$ and applying the E and M steps on multiple batches.

\spara{Implementation details.}
The implementation follows the structure depicted in the previous sections.
The vector $\eta$ is fed into a $\tanh(\cdot)$ in order to constraint its values into $[-1, +1]$,
$\theta \text{ and } \phi$ are modeled as a 2-layer GCN~\cite{kipf2016semi} using 1024 hidden units,
the social graph, one-hot encoding attributes,
and an output layer with $|C|$ components, that are then fed to a softmax and a sigmoid function, respectively for $\theta$ and $\phi$.
The latter decision is intuitive: 
echo-chamber communities compete to attract users, 
while each user could belong to multiple social communities.
For modeling, 
the implementation of $\theta$ and $\phi$ is transparent since they are normalized w.r.t. communities before the likelihood computation as mentioned at the end of Section~\ref{sec:generative-process}.
We train the overall architecture through the stochastic algorithm described above,
using Adam optimizer with default settings and one epoch.
To balance the contribution of links and propagations,
we randomly oversample the minority class between the two to achieve a balanced distribution.

\section{Experiments}
\label{sec:experiments}

\begin{table}[t]
    \caption{Results from synthetic experiments with different configurations of parameters $(s, h)$ used for generation. For each metric, we report its mean and its standard deviation across 10 experiments. Metrics indicate, respectively, the MAE between polarities for each community,
    between the original and estimated social interest of nodes in communities,
    between the original and estimated membership of nodes in communities, the correlation between original and estimated polarities for each node.
    }
    \label{tab:synthetic-exp}
    \vspace{\captionfigdist}
    \centering
    \begin{adjustbox}{max width=\columnwidth,center}
    \begin{tabular}{lllll}
        \toprule
        Input data set & $MAE(\eta, \eta^*) \downarrow$ & $MAE(\phi, \phi^*) \downarrow$ & $MAE(\theta, \theta^*) \downarrow$ & $\rho(\eta\theta, \eta^*\theta^*) \uparrow$ \\
        \midrule
        Social ($s=16$, $h=8$) & 0.27 $\pm$ 0.11  & 0.21 $\pm$ 0.00 & 0.24 $\pm$ 0.03  & 0.91 $\pm$ 0.03 \\
        Balanced ($s=8$, $h=8$) & 0.27 $\pm$ 0.10  & 0.22 $\pm$ 0.00 & 0.22 $\pm$ 0.03  & 0.93 $\pm$ 0.01 \\
        Polarized ($s=8$, $h=16$) & 0.27 $\pm$ 0.10  & 0.21 $\pm$ 0.00 & 0.19 $\pm$ 0.03  & 0.96 $\pm$ 0.03 \\
        \bottomrule
    \end{tabular}
    \end{adjustbox}
\end{table}

In this section, we empirically asses our proposal and answer the
following research questions:

\squishlist
  \item \textbf{RQ1}. Assuming that a data set is generated according to the generative process described in Section~\ref{sec:model}, is the ECD inference algorithm discussed in Section~\ref{sec:algorithms} able to estimate its original parameters? Under which conditions? (Section~\ref{sec:synthetic-experiments})
  \item \textbf{RQ2}. Do polarized communities detected in real world data sets exhibit typical features associated with echo chambers?
  (Section~\ref{sec:real-world-experiments})
  \item \textbf{RQ3}. Can our model be used to provide relevant information to auxiliary predictive tasks, such as predicting activations or individual stances?
  (Section~\ref{sec:predictive-tasks})
\squishend

\subsection{ Synthetic experiments}
\label{sec:synthetic-experiments}

In order to answer our first research question, we generate an array of data sets according to the generative model described in \Cref{sec:model}, with different combinations of hyper-parameters.

\spara{Reconstruction experiment.}
We use data sets generated with the procedure defined in \Cref{sec:model} to test whether the ECD algorithm is able to reliably infer the latent communities.
The goal here is first, to present experimental evidence that our algorithm fits the intended purpose.
Second, as with any inference procedure, it is necessary to check if a reasonably-sized amount of data is sufficient for a meaningful estimate of the latent communities with the presented algorithm, and under which conditions.

To do so, we consider the same three different settings of parameters $s, h$ from \Cref{sec:generated-networks} and visualized in \Cref{fig:examples-polarization}. %
For each parameter setting, we generate 10 data sets composed of a graph and a set of propagations each characterized by a polarity.
On each data set, we run our estimation algorithm (initialized with the values for $s, h$ used to generate the data set).
From our algorithm, we obtain an estimate for the node-community membership $\theta^*$, the community polarities $\eta^*$, and the social interest of nodes $\phi^*$.
We then measure the absolute error between the original value used to produce the data, and the estimated values obtained by our inference algorithm.
Since there is not a natural ordering in the community space, the absolute error is computed as the best result achieved through an exhaustive search in the community indexes.

We present results in Table~\ref{tab:synthetic-exp}.
We observe that our algorithm obtains a low error (between $0.19$ and $0.27$) for all parameters, and for $\phi$ and $\theta$ in particular.
Moreover, the reconstruction of the individual nodes' polarities is very precise ($\rho>0.9$).
The estimate of the node memberships $\theta$ gains reliability as the echo-chamber behavior is more apparent: this is quite expected since propagations are more informative when the network is dominated by echo chambers.

\begin{figure}
    \centering
    \hspace{-0.595cm}
    \begin{tabular}{cc}
        \includegraphics[width=.5\columnwidth]{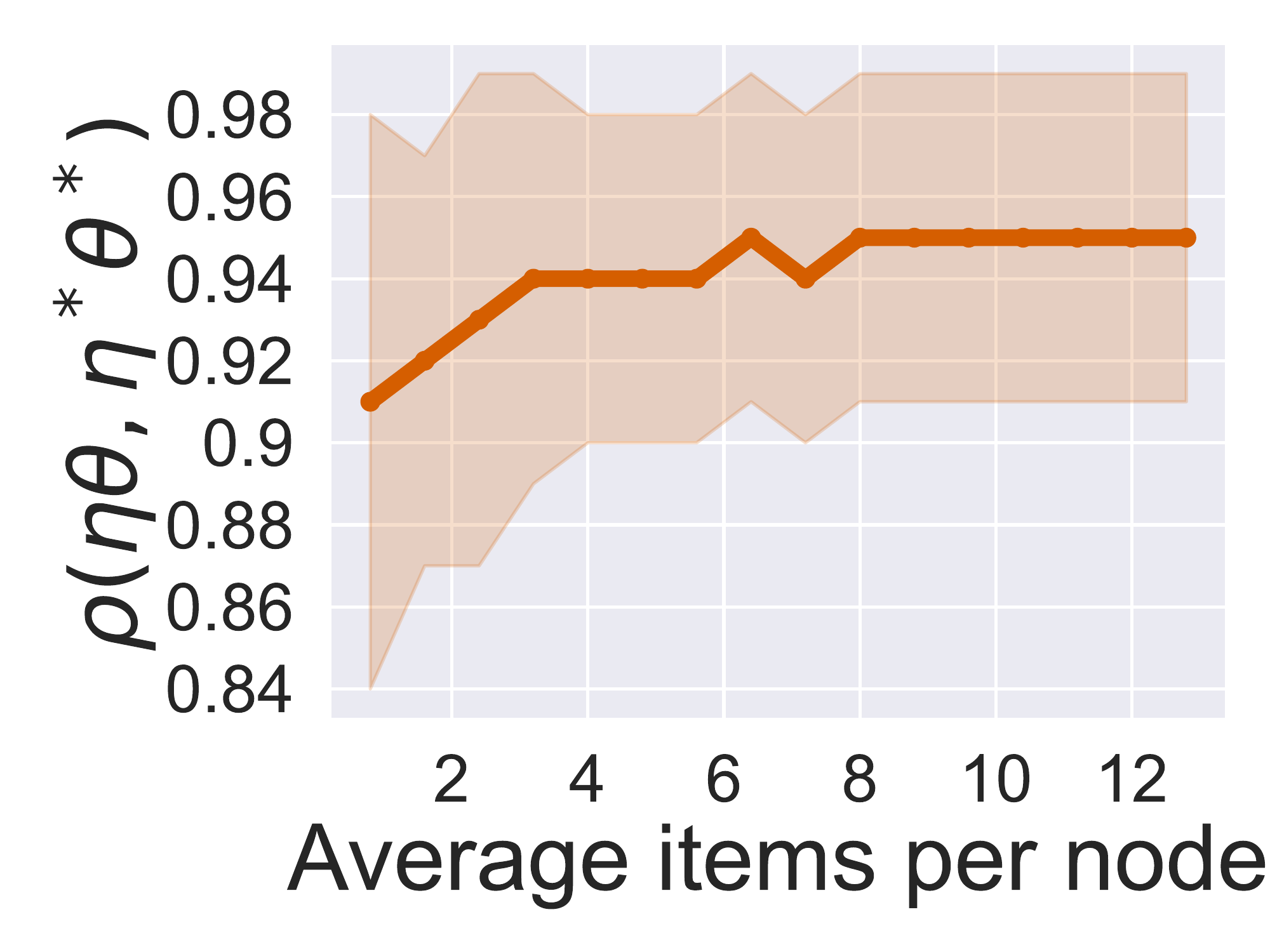}
         &
         \hspace{-0.35cm}
         \includegraphics[width=.5\columnwidth]{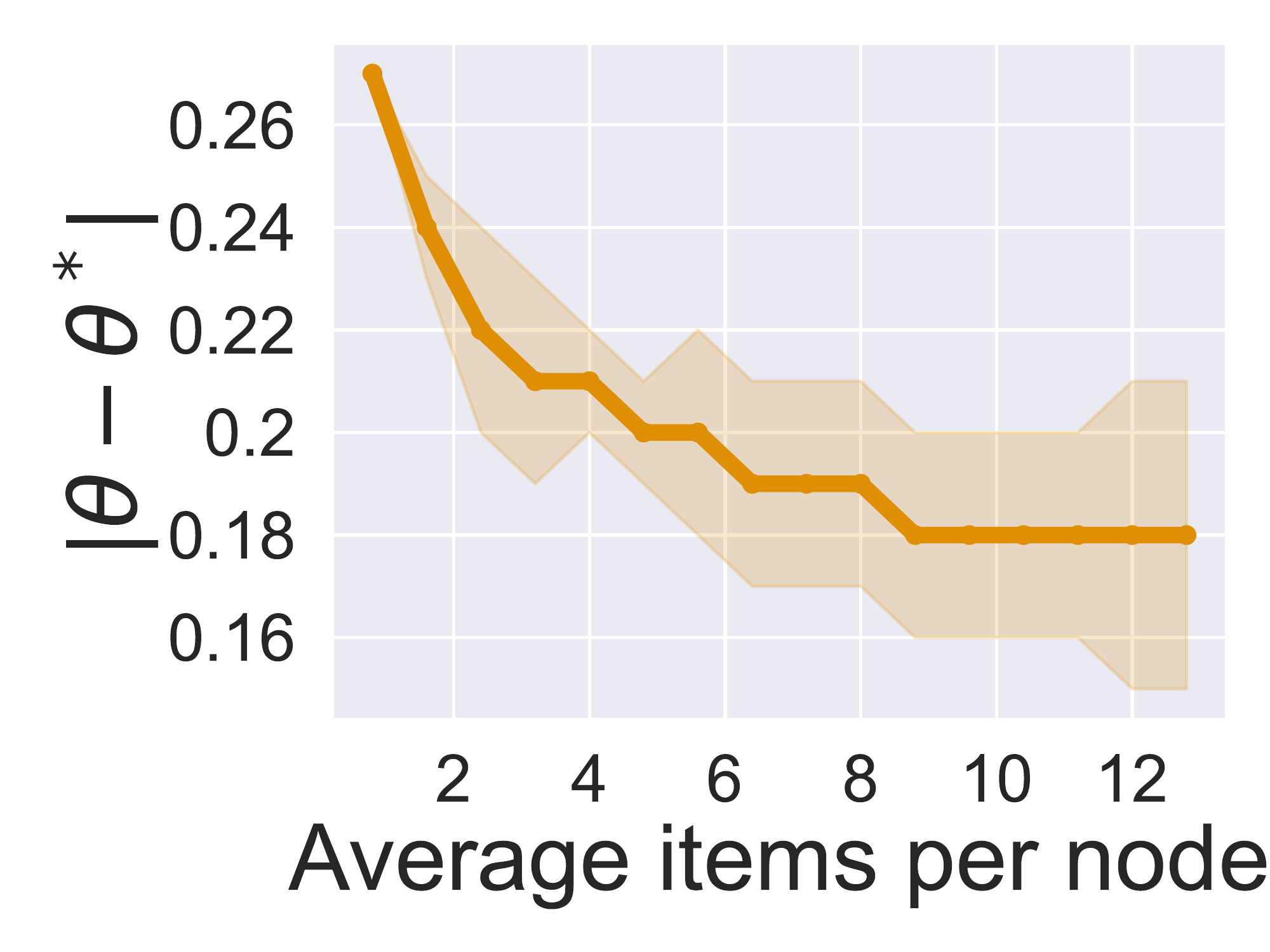}
    \end{tabular}
    \vspace{\captionfigdist}
    \caption{
    Impact of the average number of items per user on the inference of node polarities (measured by Pearson's Correlation, on the left) and community memberships (measured by MAE, on the right).
    In both cases, we observe an average number of items per user equal to 10 is sufficient to reach maximum performances, even if node polarities are recovered also with 4 items. }
    \label{fig:synthetic-nitems-tuning}
\vspace{\figtextdist}
\end{figure}

\spara{Efficiency Analysis.}
Then, we investigate the amount of data needed to reliably reconstruct the latent communities.
We do so by generating data sets with a growing number of propagations.
Specifically, we test from an average number of propagations per user of $1$ to $16$, with steps of $1$.
Using these data sets, we perform a grid of experiments by using the same setting described in Section~\ref{sec:generated-networks} ($s=8$, $h=16$).
The results are shown in Figure~\ref{fig:synthetic-nitems-tuning}.
We observe that both the estimation of node polarities and $\theta$ memberships are affected by different amounts of input data.
When looking at the membership reconstruction error, an average of 10 items per user is sufficient for the model to reach its top performance, which then saturates.
Individual node polarities, instead, are well estimated even with 4 items per user.
This analysis gives a hint at the real-world applicability of our method.

\begin{figure*}
    \centering
    \includegraphics[width=\textwidth]{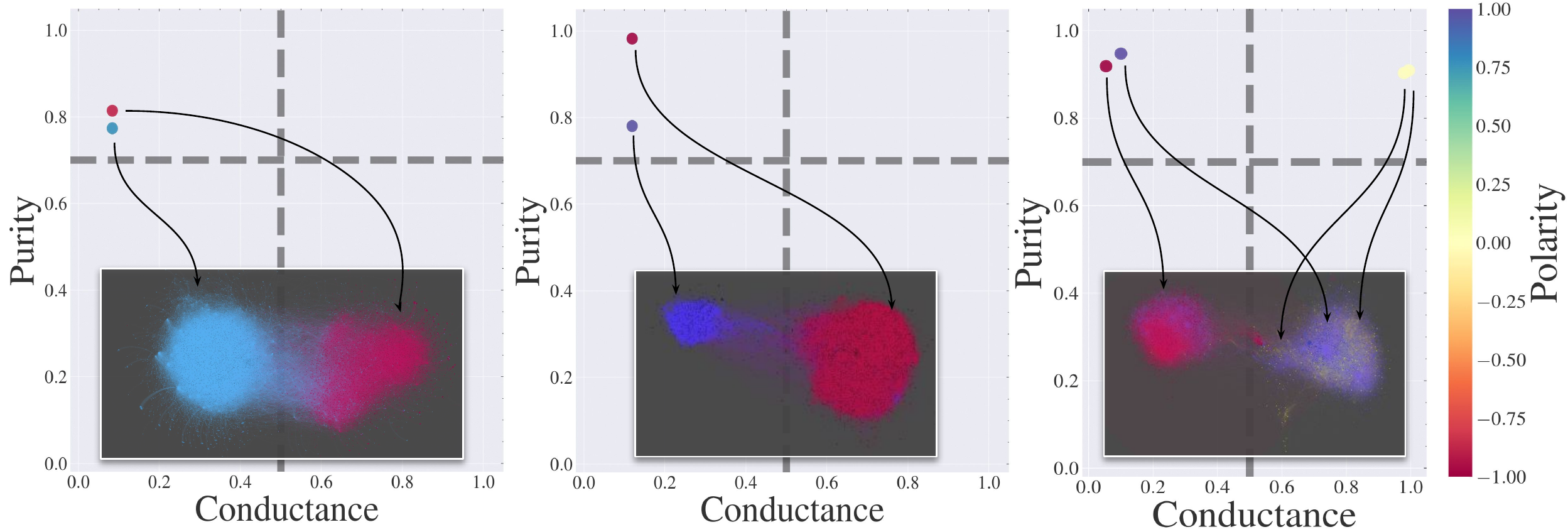}
    \vspace{\captionfigdist}
    \caption{
    Purity-Conductance plots of the communities detected by our method for three data sets: (a) \Brexit, (b) \Referendum, and (c) \VaxNoVax.
    Each dot represents a community, whose coordinates are its level of conductance (x-axis) and purity (y-axis).
    The community assignment for a user $u$ is derived from $\operatorname{argmax}_c \theta_u$, and
    the colors are associated with the $\eta$ values inferred by our model.
    For higher values of $|\eta|$ (echo chambers), we obtain communities well isolated (low conductance) and containing users of the same ideology (high purity), with a clear correspondence in the retweet networks (bottom).
    Lower absolute values of $\eta$ correspond to neutral in-between communities which correspond to news media accounts.
    Force Atlas 2 with gravity$=100$ is used in Gephi to define the layout of the network and the inferred community assignment to color the nodes.
    }
    \label{fig:real-purity-conductance}
\vspace{\figtextdist}
\end{figure*}

\subsection{ Echo chamber assessment}
\label{sec:real-world-experiments}

In order to answer our second research question, we apply ECD to three real-world data sets extracted from Twitter.

\spara{Data sets.}
Each data set is focused on a different controversial topic:
\squishlist
  \item \Brexit \cite{zhu2020neural} regards the remain-leave discourse before the 2016 UK Referendum to exit the EU. Since this data set does not include retweets, we scrape all the retweets in the period May-July 2016 that contain at least one of the first 100 most used hashtags about Brexit, for a core set of users with at least 5 tweets.
  \item \Referendum \cite{lai2018stance} is gathered during the Italian constitutional referendum in 2016.%
  \item \VaxNoVax \cite{cossard2020falling} comprises polarized discussions related to the vaccine debate in Italy in 2018.
\squishend

Summary statistics for each data set are reported in Table~\ref{tab:real-data-stats}.
Each data set exposes different features from the others, relative to the number of users and items, polarity distribution, and cascade size. %

In all data sets, we use \emph{follows} to construct the social graph $G$ and \emph{retweets} as propagations $\D$.
In cases where the two diverge (a user retweets an item that has not been shared by any of the users they follow in the graph), we insert such missing links in the social graph.
Our model also needs as input a polarity $p_i$ for each item: to compute it for the \Brexit and \Referendum data sets, we train a supervised text classifier on the labeled subset of tweets provided in the original works.
Specifically, we first subselect polarized tweets (AUC ROC 0.78 and 0.75 in 10-fold cross-validation, respectively), and then assign a polarity to each polarized tweet (AUC ROC 0.81 and 0.86).
For the \VaxNoVax data set, we use the original labeled data to train a text-based classifier that separates tweets from the two ideological sides (F1 $0.87$) and then sub-select tweets that obtain a classification score larger than $0.75$ in absolute value.

\begin{table}[t!]
    \caption{Summary statistics of real-world data sets.
    The last column refers to the ratio between the number of items with positive $\mathcal{I}^+$ and negative $\mathcal{I}^-$ polarity.
    }
    \vspace{\captionfigdist}
    \label{tab:real-data-stats}
    \centering
    \begin{adjustbox}{max width=\columnwidth,center}
    \begin{tabular}{lllll}
        \toprule
        data set & No. of users & No. of items & Cascade size & $\frac{|\mathcal{I}^+|}{|\mathcal{I}^-|}$ \\
        \midrule
        \Brexit & 7589 & 19963 & $3.6\pm 5.6$ & 1.37 \\
        \Referendum  & 2879 & 40344 & $5.6\pm 9.9$ & 0.18 \\
        \VaxNoVax & 14315 & 21312 & $7.6\pm 28.9$ & 0.80 \\
        \bottomrule
    \end{tabular}
    \end{adjustbox}
\vspace{-3mm}
\end{table}

\spara{Experimental protocol.}
We next apply the ECD algorithm to these data sets.
We chose social prior size $s=8$ and echo-chamber prior size $h=16$, since these data sets are collected around polarizing topics, where we expect to find a configuration similar to Figure~\ref{fig:examples-polarized}.
We set $K=8$ as the number of communities; if the data set can be explained by a fewer number of communities, our method simply assigns them a near-zero membership.
Thus, in the following, we will consider only non-empty communities.
We then adopt the evaluation method proposed by \citet{morini2021toward}: analyzing each community in terms of its \emph{conductance} ---i.e., how closely-knitted is the community with the rest of the graph---and its \emph{purity}---i.e., the ratio of users with the same ideological alignment, measured as the average polarity of the tweets they reshare.
Morini~et~al. identify a low conductance and a high purity as typical properties of echo chambers.
ECD training time takes $\sim$2, 6, and 120 minutes on the three dataset \Brexit, \Referendum, and \VaxNoVax.
In light of Table~\ref{tab:real-data-stats}, cascade size is a determinant factor for scalability.
This is intuitive since we model cascades through all pairs of users $\p(u,v,i)$ that interacts with item $i$.

\spara{Results.}
We report our results in Figure~\ref{fig:real-purity-conductance}.
We observe that all the echo chambers detected by our method (i.e., the communities with a high value of $|\eta|$) indeed display typical echo-chamber traits.
Specifically, in the case of \Brexit and \Referendum (Figures~\ref{fig:real-purity-conductance}a, \ref{fig:real-purity-conductance}b), we obtain two echo-chamber communities with high purity and low conductance.
On \VaxNoVax, besides the two echo-chamber communities we also obtain two social communities ($|\eta| \sim 0$).
From an empirical analysis, one of these social communities contains all authoritative news sources (e.g., SkyTG24 and AdKronos), while the other contains users who are arguably pro-vax, but without significant pro-vax propagations.
Indeed, both social communities have very high conductance, thus missing the segregation exhibited by typical echo chambers.

\subsection{Predictive tasks}
\label{sec:predictive-tasks}
The latent communities and echo chambers discovered by the ECD model provide valuable information to describe social media users.
Such information could be therefore useful for other applicative predictive tasks.
To assess the significance of the produced communities in such tasks, we study two typical prediction problems: \emph{graph-based stance detection} and \emph{next-activation prediction}.

\spara{Graph-based stance detection.}
In the first task, we wish to assign an individual polarity to each node in the network.
To evaluate our performance, we manually label a set of $\sim$100 users for each data set.
Then, we apply our model, excluding their activations from the training set~(our model sees their social links).
Finally, we assign to each of them the polarity as the weighted average of the polarities of the communities they belong (i.e., $\eta^* \cdot \theta^*$).

We compare its results to the following baselines:
\squishlist
    \item \emph{1-Hop Average}: given a user $u$ we compute their stance as the average polarity of the propagations of the users that $u$ follows.
    A similar method was proposed by \citet{barbera2015birds}.
    \item \emph{node2vec} \cite{grover2016node2vec}: we embed the social graph $G$ using the embedding dimension $K=128$; then, we train a logistic regression using these embeddings. %
    Since this method is supervised, we test it through a leave-one-out cross-validation procedure w.r.t. our set of manually-labeled users; we then report the average.
    \item \emph{GCN} \cite{kipf2016semi}: we adopt a 2-layer Graph Convolutional Neural Network, using as node features $x_u$ a one-hot encoding of the $|I|$ propagations, i.e. $x_u[i]=1$ if $u\in \D_i$, and 0 otherwise.
    Since this method is also supervised, we again adopt leave-one-out to test its performance.
\squishend

Results are reported in Table~\ref{tab:stance-detection}.
We use ROC-AUC as a standard metric to compare the superiority of different models.
On all three data sets, our method significantly outperforms the 1-Hop Average baseline.
Node2Vec and GCN perform substantially worse than our method, except on \VaxNoVax, where all methods achieve good results.
In practice, although propagations or the social graph are valuable sources, ECD is the only model that can efficiently combine information coming from both.%

\spara{Next-activation prediction.}
The second predictive task we test is predicting future propagations.
In order to evaluate our approach, for each propagation, we split the set of nodes that activated on it into training and test.
That is, a fraction (to be determined later) of the activated nodes is not visible during training.
Then, we use only the training activations to estimate our model parameters.
To approximate the probability of a node $u$ activating on an item $i$, we consider the maximum probability of propagation from each of the other activated nodes $D_i$;
then, following our model, each probability is computed marginalizing over each community $c$, thus obtaining
\begin{equation*}
P(u|\D_i) = \max\left\{\sum_{c} \pi_f(c) \cdot P( \p(u,v,p_i)|c) \; \middle| \;\forall v \in \D_i\right\}.
\label{eq:next-act}
\end{equation*}

Using this probability as a prediction score, we evaluate the performance as a binary classification task where, given a pair $(u, i)$, the model predicts whether user $u$ will activate on item $i$ or not.
Hence, we use ROC AUC to measure prediction quality.
We apply this procedure for each of the three data sets introduced in the previous section.
On each data set, we test different fractions for the train-test split, expressed as the percentage of masked activations during training.
Since we treat the next-activation task as a binary classification problem,
all pairs $(u, i)$ s.t. $u \notin \D_i$ are attached to the test set as negative instances.

\begin{table}
    \caption{
    ROC-AUC on the stance-detection task for our approach (ECD) and different graph-based supervised and unsupervised baselines (see text).
    ROC-AUC scores of supervised baselines are measured as average on leave-one-out cross validation.
    }
    \vspace{\captionfigdist}
    \label{tab:stance-detection}
    \centering
    \begin{adjustbox}{max width=\columnwidth,center}
    \begin{tabular}{lllll}
        \toprule
        Supervised & Method & \multicolumn{3}{c}{ROC-AUC}  \\
         & & Brexit & VaxNoVax & Referendum \\
        \midrule
         & ECD & 0.98 & 0.97 & 0.91 \\
         & 1-Hop Average & 0.47 & 0.85 & 0.49 \\
        \checkmark & Node2Vec+LR & 0.85 & $1.00$ & 0.75 \\
        \checkmark & GCN & 0.92 & 0.94 & 0.87 \\
        \bottomrule
    \end{tabular}
    \end{adjustbox}
\vspace{-3mm}
\end{table}

To benchmark the performance of our method, we compare its results with two heuristics.
The first one, dubbed $Most Pop$, gives higher probability to the most active users:
\begin{equation*}
  \label{eq:most-pop}
  \operatorname{MostPop}(u, \D_i) = \frac{\sum_{j \in \I} \mathbbm{1}(u \in \D_j)}{\sum_{j \in \I} \sum_{v \in V} \mathbbm{1}(v \in \D_j)}
\end{equation*}
while the second, ${Most Pop}^*$, takes into account the item polarity by assigning higher weight to those users
activating on items with similar polarity:
\begin{equation*}
  \label{eq:most-pop-star}
  \operatorname{MostPop}^*(u, \D_i) = \frac{\sum_{j \in \I, sign(p_i)=sign(p_j)} \mathbbm{1}(u \in \D_j)}{\sum_{j \in \I, sign(p_i)=sign(p_j)} \sum_{v \in V} \mathbbm{1}(v \in \D_j)}
\end{equation*}

\begin{figure}
\Large Next-Activation prediction
\begin{tabular}{cc}
  \includegraphics[width=.45\columnwidth]{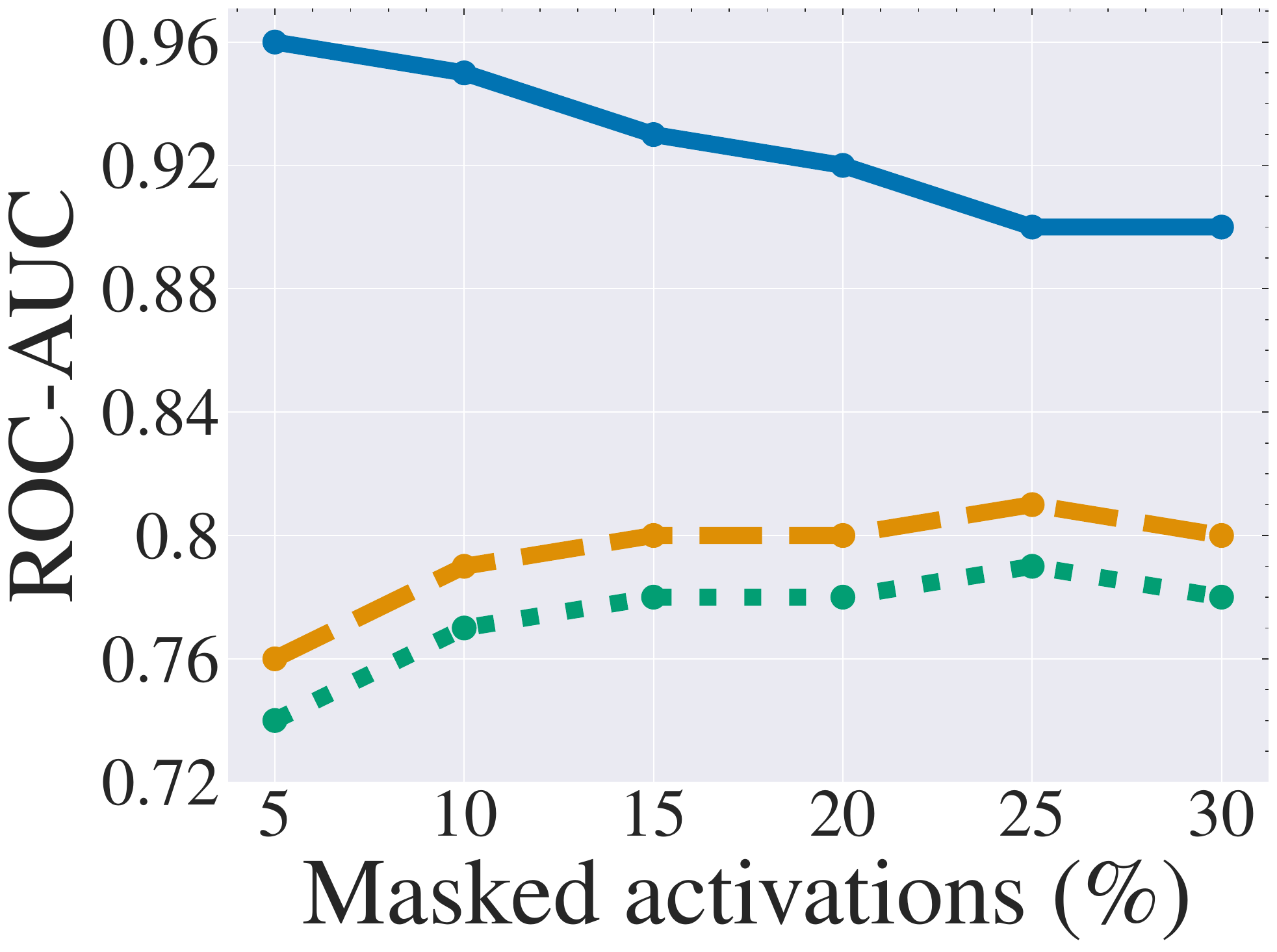} &   \includegraphics[width=.45\columnwidth]{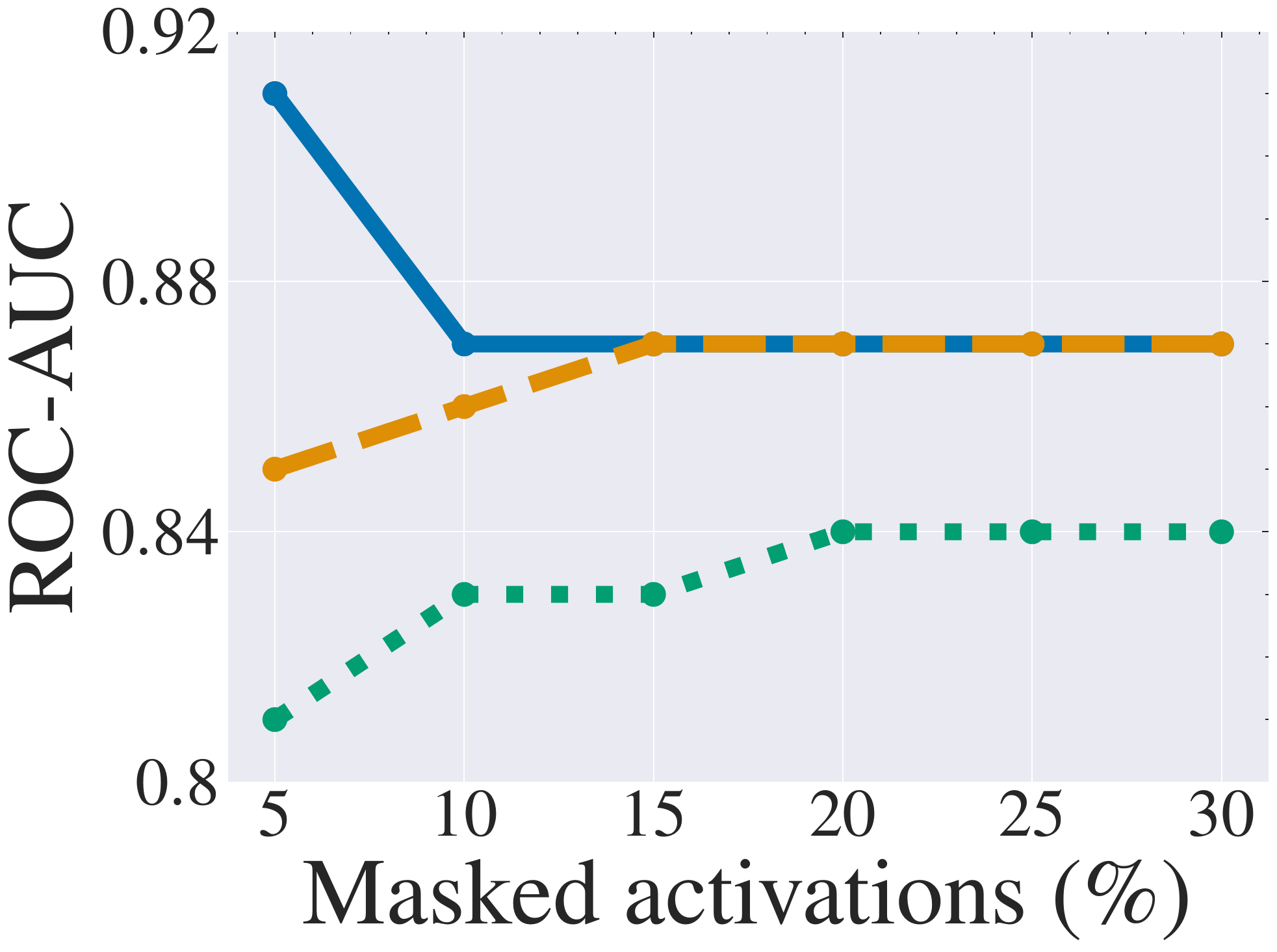} \\
\hspace{0.45cm} (a) Brexit & \hspace{0.45cm} (b) Referendum \\[6pt]
\end{tabular}
\begin{tabular}{p{0.7\columnwidth} p{0.3\columnwidth}}
  \vspace{-15.25pt} \hspace{1cm} \includegraphics[width=0.45\columnwidth]{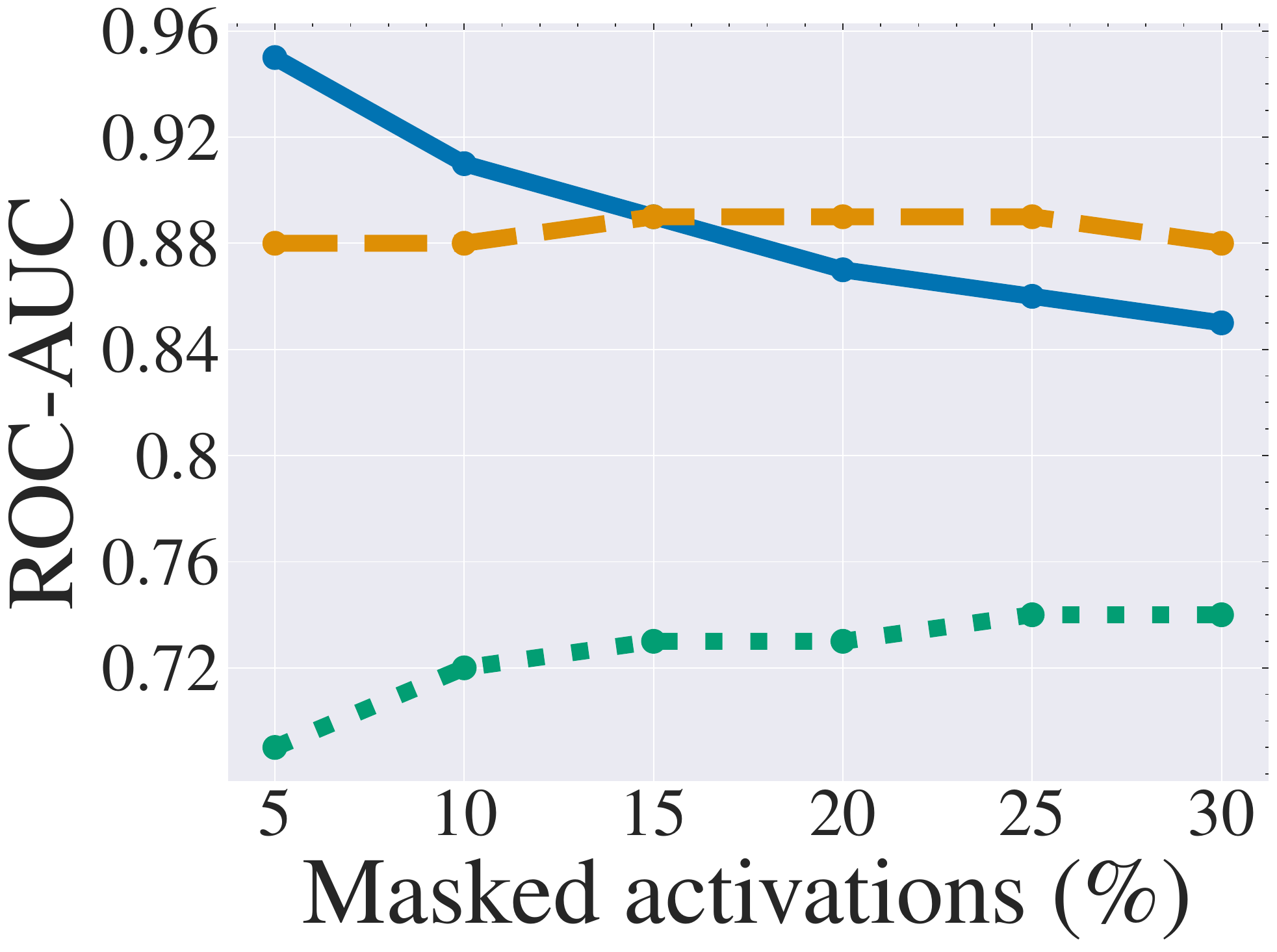} &
  \vspace{-15.25pt} \hspace{-1.2cm} \includegraphics[width=0.3\columnwidth]{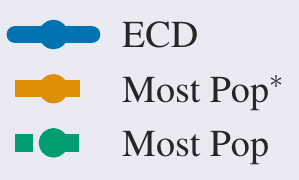} \\
  \hspace{2.25cm} (c) VaxNoVax & {}
\end{tabular}
 \\[6pt]
 \vspace{\captionfigdist}
\caption{
Performance of our ECD model and baselines on the task of next-activation prediction, for our three data sets, as a function of the training set size.
}
\label{fig:next-act}
\vspace{\figtextdist}
\end{figure}

Results are shown in Figure~\ref{fig:next-act}.
With a 90\%-10\% train-test split, we report an AUC ROC of around $0.9$ for all three data sets---substantially better than the tested baselines.
Moreover, the performance of our method degrades gracefully when the train set size decreases: we do not observe sudden variations in these curves.
This result also suggests that the model is not impacted by different sampling choices for the training set.

\smallskip
\noindent \textbf{Reproducibility:} our code and data are available at\\
\url{https://github.com/mminici/Echo-Chamber-Detection.git}

\section{Conclusions and Future Work}
\label{sec:conclusions}
In this work, we fill the gap between modeling and data analysis approaches when studying echo chambers in social networks.
We propose a gradient-based inference algorithm derived from a probabilistic model, which
implements realistic assumptions on echo chambers, distinguishing them from other types of communities, and can be used to generate polarized networks and propagations.

Our solution inherits its explainability from this principled generative approach.
This approach allows us to formalize the common intuition of a deep entanglement between the observed propagation patterns of polarized contents, and the latent association between users and communities.

The experimental analysis confirms that our algorithm successfully detects echo chambers exhibitng their typical traits of connectivity and opinion homogeneity.
Comparisons against state-of-the-art baselines on auxiliary prediction tasks, such as stance detection and next-activation prediction, show the good performance of our algorithm for such tasks in cold-start settings.
Also, experiments show that the algorithm is efficient in terms of the number of input propagations needed, and robust with respect to missing data.

Our approach relies on minimal and realistic assumptions that define the perimeter of its effectiveness, and allows for possible extensions.
For instance, we consider one specific type of interaction that reflects endorsement, and neglects all the possible nuances in the users' debates (e.g., \emph{replies} on Twitter could be antagonizing).
Furthermore, as presented in Section~\ref{sec:synthetic-experiments}, the results of our method improve with the polarization of the input.
However, it would be straightforward to extend our model by introducing a form of hindered propagation for neutral content in non-echo-chamber communities; such an extension would leave our algorithm almost identical.

As in every experimental study, our empirical validation is limited by the available data.
For instance, we use only one type of social network, i.e. Twitter.
Nonetheless, we hypothesize that our framework would suit also other social media platforms, as long as they allow the existence of a social graph and propagations.

Our work focuses on a given single ideological axis; however, learning the interplay of different axes has been proved successful in the literature~\cite{monti2021learning}: it would be worth devising an extension of our model able to deal with multiple ideological axes.

Another interesting direction for future investigation, would be to model the temporal aspects of propagations that, for simplicity, we do not consider here.
On the one hand, propagations naturally happen over time, and their speed could provide further characterization of echo chambers \cite{Manco2019}.
On the other hand, also polarities do change over time: such an extension could offer improvements on the difficult task of learning opinion dynamics from data~\cite{monti2020learning}.

Finally, we consider the polarity of items (i.e., tweets) as part of the input, since they can be easily obtained from natural language processing techniques.
However, it would be interesting to integrate such approaches with our model, by considering the items' polarities as a latent variable, that can be estimated by looking at their propagations as well as their content.

\section*{Acknowledgments}
Marco Minici and Giuseppe Manco acknowledge partial support by the EU H2020 ICT48 project ``HumanE-AI-Net" under contract \#952026. Federico Cinus acknowledges partial support by SoBigData++ through the Transnational Access project.

\bibliographystyle{ACM-Reference-Format}
\bibliography{references}

\clearpage
\renewcommand{\thesubsection}{\Alph{subsection}}

\end{document}